\documentclass[12pt,amsmath,amssymb,]{revtex4}
\usepackage{graphics}
\usepackage{amsfonts}
\usepackage{amsmath}
\usepackage{epsfig}
\usepackage{epsf}
\usepackage{graphicx}
\usepackage{dcolumn}
\usepackage{bm}
\usepackage{color}
\usepackage{array}
\usepackage{mathrsfs}
\usepackage{amsmath}
\usepackage{braket}
\usepackage{indentfirst}
\usepackage{extarrows}
\usepackage{harpoon}

\newcommand {\sla}[1]{ #1 \!\!\!/}

\def\ds{\displaystyle}

\begin{document}

\title{ Two-photon-exchange effects in the unpolarized $\mu p$ scattering within the hadronic model}

\author{
Hai-Qing Zhou$^{1,2}$\protect\footnotemark[1]\protect\footnotetext[1]{E-mail: zhouhq@seu.edu.cn}\\
$^1$Department of Physics,
Southeast University, NanJing 211189, China\\
$^2$State Key Laboratory of Theoretical Physics, Institute of Theoretical Physics,\\
Chinese Academy of Sciences, Beijing\ 100190,\ P. R. China }
\date{\today}

\begin{abstract}
In this work, the two-photon-exchange (TPE) effects in the unpolarized $\mu p$ scattering are discussed within the hadronic model where the intermediate states $N,\Delta$ and $\sigma$ are considered. The contribution from the $N$ intermediate is close to the results given by Ref. \cite{Tomalak2014} at the small $Q$ and there is a sizeable difference when $Q>0.25$GeV (where $Q^2$ is the four momentum transfer). The contributions from the $\Delta$ and the $\sigma$ intermediate states are much smaller than that from the $N$ intermediate at the small $Q$. In the kinematic region with $k_i\subseteq [0.01,0.3]$ GeV  and $Q \leq0.4$GeV (where $k_i$ is the three momentum of initial muon at Lab frame), a naive expression for the TPE contributions is given, which can be  used directly for other analysis.

\end{abstract}

\maketitle

\section{Introduction}
The two-photon-exchange (TPE) effects in the elastic $ep$ scattering have been widely studied (see the recent review paper \cite{Carlson07,Arrington2011}) after 2000 to explain the discrepancy between the measurements of $R =\mu G_E/G_M$ (with $G_{E,M}$ the electromagnetic form factors of proton) by the Reosenbluth method \cite{Andivahis1994,Walker1994} and the polarized method \cite{Jones2000,Gayou2002}. After the arising of the puzzle of proton \cite{Pohl2010,Antognini2013-Science}, the TPE effects in the $\mu p$ system also abstract many interestings \cite{two-pion-exchange-to-energy-spectrum,Alarcon2014,one-pion-energy-spectrum-1,one-pion-energy-spectrum-2}. The coming experiment MUSE \cite{MUSE} proposes the measurement of the electromagnetic form factors of proton by the elastic unpolarized $\mu p$ scattering at the small momentum transfer and the aim of the precise extraction of the form factors calls for the careful consideration on the TPE effects.

In the literature, many methods have been applied to estimate the TPE effects in the $ep$ and $\mu p$ scattering, for example, the hadronic model
\cite{Blunden03,Kondra05,Blunden05,Dian-Yong-Chen2013,Tomalak2014}, GPD method \cite{Chen04,Afana05}, phenomenological parametrizations \cite{Chen07,BK07}, dispersion relation approach \cite{BK06,BK08,BK11,BK12,BK14}, pQCD calculations \cite{BK09,Kivel09} and SCEF method \cite{TPE-SCEF}. Among these methods, the hadronic model is usually used at the small and medium momentum transfer.  By this method, the TPE contribution in the $\mu p$ scattering from the intermediate $N$ was estimated in Ref. \cite{Dian-Yong-Chen2013,Tomalak2014}, and recently the contribution from the $\sigma$ meson exchange in the $t$-channel was also discussed in \cite{Koshchii2016}. In this work, we give an estimation of the TPE effects in the $\mu p$ system from the intermediate state $\Delta$, and the contributions from the $N$ and the $\sigma$ intermediate states are also discussed. And furthermore, we give a naive formula to express these contributions, which can be  used directly for other analysis. In Sec.II, we give a brief introduction of the model, in Set. III, we list the input parameters we used, in Sec IV we present the numerical results and at last we give a discussion and a short summary.

\section{Basic Formula}

\begin{figure}[htbp]
\center{\epsfxsize 2.0 truein\epsfbox{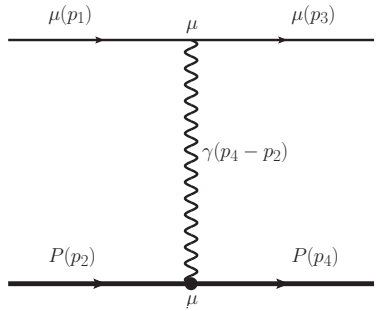}}
\caption{One-photon-exchange diagram for $\mu p$ scattering in Feynman gauge.}
\label{figure:mup-OPE}
\end{figure}

In Feynman gauge, the amplitude for the $\mu p$ scattering in the Bonn approximation showed in Fig. \ref{figure:mup-OPE} can be expressed as
\begin{eqnarray}
i\mathcal{M}^{1\gamma}_{\mu p} &=&  \overline{u}(p_3,m_\mu)(-ie\gamma_{\mu})u(p_1,m_\mu)\overline{u}(p_4,m_N)\Gamma_{\gamma NN}^{\mu}u(p_2,m_N)S_\gamma(q),
\label{eq:M-OPE}
\end{eqnarray}
with $p_1,p_3$ the momentums of the incoming and outgoing  muons, $p_2,p_4$ the momentums of the incoming and outgoing protons, $m_\mu,m_N$ the masses of muon and proton, $e=-|e|$, $\Gamma_{\gamma NN}^{\mu}$ the effective vertex for $\gamma NN$ interaction, $S_\gamma(q) = \frac{-i}{q^2+i\epsilon}$ and $q\equiv p_4-p_2$.

The TPE amplitude in the $\mu p$ scattering generally can be expressed as
\begin{eqnarray}
i\mathcal{M} ^{2\gamma}_{\mu p} &=&  \int \frac{d^4\overline{k}}{(2\pi)^4} L^{\mu\nu}_{\mu p}H_{\mu p,\mu\nu},
\label{eq:M-TPE}
\end{eqnarray}
where $L^{\mu\nu}_{\mu p}$ is the amplitude for  the double virtual Compoton scattering of muon which can be written down explicitly, and $H_{\mu p,\mu\nu}$ is the amplitude for the double virtual Compton scattering of proton. Due to the non-perturbative properties of QCD, the explicit expression for $H_{\mu p,\mu\nu}$ in all the kinematical region is unknown. In the very low momentum region, this amplitude can be estimated by the chiral perturbative theory, and in the deep virtual region, it can be estimated by the GPD, pQCD or SCEF methods. In the medium momentum transfer region, the estimation based on the hadronic level has to be applied.

At hadronic level, we separate the contributions to the double virtual Compton scattering into four kinds as the $s$-channel, $u$-channel, $t$-channel and the other contributions.

In this work, we  limit our discussion in the former three kinds of contributions, and in the $s$- and $u$- channels, we only considered the $N$ and the $\Delta$ intermediate states, and in the $t$- channel, we only consider the $\sigma$ meson exchange.

\subsection{TPE contributions from $N,\Delta$ intermediate states in $s,u$-channels}

\begin{figure}[htbp]
\center{\epsfxsize 4.0 truein\epsfbox{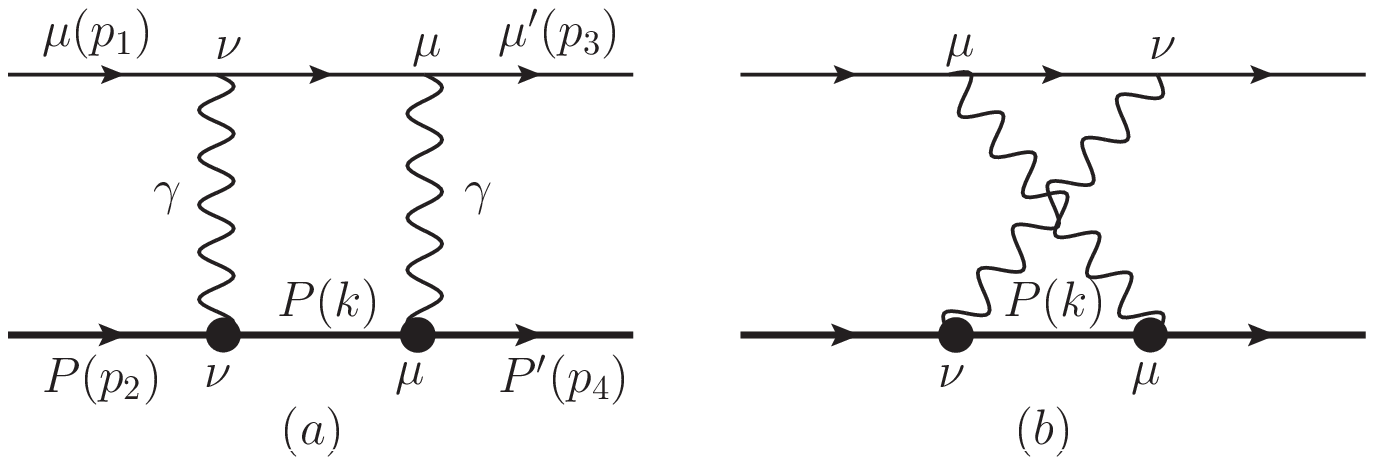}}
\center{\epsfxsize 4.0 truein\epsfbox{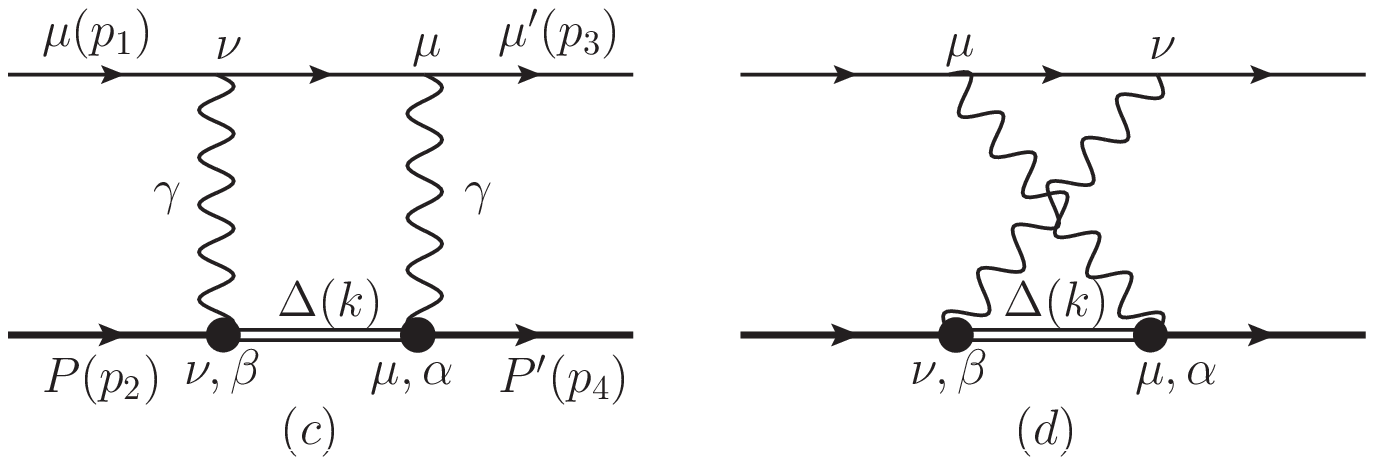}}
\caption{TPE contributions to $\mu p$ scattering from the $s,u$-channels due to the $N,\Delta$ intermediate states in Feynman gauge.}
\label{figure:TPE-ND}
\end{figure}

At hadronic level, the diagrams for the TPE amplitudes in the $s$- and $u$ -channels of the $\mu p$ scattering are showed as Fig.\ref{figure:TPE-ND}, where the intermediate states are proton and $\Delta$.

In our estimation, we take the corresponding effective vertexes as
\begin{eqnarray}
\Gamma_{\gamma NN}^{\mu}(\overline{q}) &=& ie\bigg\{F_1(\overline{q}^2)\gamma^{\mu}+i\sigma^{\mu\nu}\frac{F_2(\overline{q}^2)}{2m_N}q\overline{}_{\nu}\bigg\}, \nonumber\\
\Gamma_{\gamma \Delta \rightarrow N}^{\mu \alpha}(\overline{p},\overline{q})
 &=& -i \sqrt{\frac{2}{3}}{\ds \frac{e}{2 m_{\Delta}^2}} \bigg\{
g_1 F^{(1)}_\Delta(\overline{q}^2) [ g^{\mu \alpha} \sla{\overline{p}}~\sla{\overline{q}} -
\overline{p}^\mu \gamma^\alpha \sla{\overline{q}}
- \gamma^\mu \gamma^\alpha \overline{p} \cdot \overline{q} + \gamma^\mu \sla{\overline{p}}~\overline{q}^\alpha ] \nonumber\\ & &+ g_2 F^{(2)}_\Delta(\overline{q}^2)\left[\, \overline{p}^\mu \overline{q}^\alpha - g^{\mu \alpha} \overline{p} \cdot \overline{q}\, \right] \nonumber\\ & &
+ (g_3/M_{\Delta})F^{(3)}_\Delta(\overline{q}^2) [ \overline{q}^2 (\overline{p}^\mu
\gamma^\alpha - g^{\mu \alpha} \sla{\overline{p}})
 + \overline{q}^\mu (\overline{q}^\alpha \sla{\overline{p}} -
\gamma^\alpha \overline{p} \cdot \overline{q} ) ] \bigg\} \gamma_5,
\nonumber \\
\Gamma_{\gamma N \rightarrow \Delta}^{\nu \beta}(\overline{p}~\overline{,q})
&=& -i\sqrt{\frac{2}{3}} {\ds \frac{e}{2 m_{\Delta}^2}} \gamma_5\bigg\{
g_1F^{(1)}_\Delta(\overline{q}^2) [ g^{\nu \beta} \sla{\overline{q}}\sla{\overline{p}}
-p^\nu \sla{q}\gamma^\beta
- \gamma^\beta \gamma^\nu \overline{ p} \cdot \overline{q} + \sla{\overline{p}}\gamma^\nu  \overline{q}^\beta ] \nonumber\\ &&
+ g_2F^{(2)}_\Delta(\overline{q}^2) [ \overline{p}^\nu \overline{q}^\beta - g^{\nu \beta} \overline{p} \cdot\overline{ q} ] \nonumber\\ &&
- (g_3/m_{\Delta})F^{(3)}_\Delta(\overline{q}^2) [ \overline{q}^2 (\overline{p}^\nu
\gamma^\beta - g^{\nu \beta} \sla{\overline{p}})
+ \overline{q}^\nu (\overline{q}^\beta \sla{\overline{p}} -
\gamma^\beta \overline{p} \cdot \overline{q} ) ] \bigg\},
\label{eq: Vertex-ND-gamma}
\end{eqnarray}
with $\overline{q},\overline{p}$ the momentums of the incoming photon and proton or $\Delta$, $m_\Delta$ the mass of $\Delta$ and $F_{1,2},F_\Delta^{(1,2,3)}$ the corresponding form factors.

By these effective vertexes, the corresponding amplitudes in Feynman gauge can be written down explicitly as
\begin{eqnarray}
i\mathcal{M}^{(a)}_{\mu p}&=&\int\frac{d^4k}{(2\pi)^4}\overline{u}(p_3,m_\mu)(-ie\gamma_{\mu}) S_\mu(p_1+p_2-k)(-ie\gamma_{\nu})u(p_1,m_\mu) S_\gamma(p_4-k)S_\gamma(k-p_2)  \nonumber \\
& &  \times\overline{u}(p_4,m_N)\Gamma^{\mu}_{\gamma N N}(p_{4}-k)S_p(k)  \Gamma^{\nu}_{\gamma N N}(k-p_{2})u(p_2,m_N),\nonumber \\
i\mathcal{M}^{(b)}_{\mu p}&=&\int\frac{d^4k}{(2\pi)^4}\overline{u}(p_3,m_\mu)(-ie\gamma_{\nu}) S_\mu(p_1-p_4+k)(-ie\gamma_{\mu})u(p_1,m_\mu) S_\gamma(p_4-k)S_\gamma(k-p_2)  \nonumber \\
& &  \times\overline{u}(p_4,m_N)\Gamma^{\mu}_{\gamma N N}(p_{4}-k)S_p(k)  \Gamma^{\nu}_{\gamma N N}(k-p_{2})u(p_2,m_N),\nonumber \\
i\mathcal{M}^{(c)}_{\mu p}&=&\int\frac{d^4k}{(2\pi)^4}\overline{u}(p_3,m_\mu)(-ie\gamma_{\mu}) S_\mu(p_1+p_2-k)(-ie\gamma_{\nu})u(p_1,m_\mu) S_\gamma(p_4-k)S_\gamma(k-p_2)  \nonumber \\
& &  \times\overline{u}(p_4,m_N)\Gamma^{\mu\alpha}_{\gamma\Delta\rightarrow N}(k,p_{4}-k) S_{\Delta,\alpha\beta}
 \Gamma^{\nu\beta}_{\gamma N\rightarrow \Delta}(k,k-p_{2})u(p_2,m_N), \nonumber \\
 i\mathcal{M}^{(d)}_{\mu p}&=&\int\frac{d^4k}{(2\pi)^4}\overline{u}(p_3,m_\mu)(-ie\gamma_{\nu}) S_\mu(p_1-p_4+k)(-ie\gamma_{\mu})u(p_1,m_\mu) S_\gamma(p_4-k)S_\gamma(k-p_2)  \nonumber \\
& &  \times\overline{u}(p_4,m_N)\Gamma^{\mu\alpha}_{\gamma\Delta\rightarrow N}(k,p_{4}-k) S_{\Delta,\alpha\beta}
 \Gamma^{\nu\beta}_{\gamma N\rightarrow \Delta}(k,k-p_{2})u(p_2,m_N),
 \label{eq:box-amp}
\end{eqnarray}
with
\begin{eqnarray}
S_\mu(\overline{k}) &=& \frac{i(\sla{\overline{k}}+m_\mu)}{\overline{k}^2-m_\mu^2+i\epsilon}, \nonumber \\
S_N(\overline{k}) &=& \frac{i(\sla{\overline{k}}+m_N)}{\overline{k}^2-m_N^2+i\epsilon}, \nonumber \\
S_{\Delta,\alpha\beta}(\overline{k})&=&\frac{-i(\sla{\overline{k}}+m_{\Delta})}{\overline{k}^2-m_{\Delta}^2+i\varepsilon}P_{\alpha\beta}^{3/2}(\overline{k}), \nonumber \\
P_{\alpha\beta}^{3/2}(\overline{k}) &=& g_{\alpha\beta}-\frac{\gamma_{\alpha}\gamma_{\beta}}{3}
-\frac{(\sla{\overline{k}}\gamma_{\alpha}\overline{k}_{\beta}+\overline{k}_{\alpha}\gamma_{\beta}\sla{\overline{k}})}{3\overline{k}^2}.
\end{eqnarray}

\subsection{TPE contribution from $\sigma$ intermediate state in $t$-channel}

The meason exchange effect in the lepton proton scattering was studied firstly in the $ep$ scattering case in Ref. \cite{zhouhq2014}, where it was pointed out that by the current precise experimental data sets at $Q^2\equiv-q^2\sim2.5$GeV$^2$ \cite{Qattan2005,Qattan2006,Ex-polarized-Meziane-2011}, the contribution from the  $2^{++}$ meson exchange should be considered. In the $ep$ scattering case, when $Q^2\gg m_e^2$ and the approximation $m_e=0$ is taken, the contributions from the $0^{-+}$ and $0^{++}$ mesons exchange are zero due to the zero mass $m_e$. While in the $\mu p$ system, these contributions maybe play their roles. The contribution from the $0^{-+}$ meson (pion) in the Lamb shift of the $\mu p$ system has been discussed in Ref. \cite{one-pion-energy-spectrum-1,one-pion-energy-spectrum-2} and is found to be very small due to the chiral anomaly. And recently the contributions from the $\sigma$ meson in the $\mu p$ scattering and the $\mu p$ bound state were discussed in Ref. \cite{Koshchii2016} and Ref. \cite{zhouhq2016}. In Ref. \cite{Koshchii2016}, the contribution from the $\sigma$ meson exchange is calculated based on the direct effective coupling of $\sigma \gamma \gamma$ with a coupling constant $g_{\sigma \gamma\gamma}$ (for real photon case). And such $q_\sigma^2$ independent effective coupling constant $g_{\sigma\gamma\gamma}$ (where $q_{\sigma}$ is the four momentum of $\sigma$) is determined from the decay width $\Gamma_{\sigma \rightarrow 2\gamma}$. This is not a good way due to two reasons: (1) the sign of the effective coupling $g_{\sigma \gamma\gamma}$ can not be determined just from the decay width $\Gamma_{\sigma \rightarrow 2\gamma}$, (2) the effective coupling $g_{\sigma \gamma \gamma}$ in the space like is very different with that in the time like region, for example, it is real in the space like region while it is complex in the time like region when $q_{\sigma}^2>4m_\pi^2$.  In Ref. \cite{zhouhq2016}, the contribution from the $\sigma$ meson exchange is estimated from the $\sigma \pi\pi $ and $\sigma NN$ couplings by the loop effects where the $q_\sigma^2$ dependence of the effective coupling is included. In this work, we follow the method used in Ref. \cite{zhouhq2016} and take the following effective vertexes to estimate the TPE contribution in the $\mu p$ scattering due to the $\sigma$ meson exchange,
\begin{eqnarray}
\widetilde{\Gamma}_{\sigma NN} &=& -ig_{\sigma NN},  \nonumber \\
\widetilde{\Gamma}_{\sigma\pi\pi} &=& -ig_{\sigma\pi\pi}, \nonumber \\
\widetilde{\Gamma}_{\gamma\pi\pi}^{\mu} &=&-ie(p_1^\mu+p_2^\mu)F_\pi(\overline{q}^2),\nonumber \\
\widetilde{\Gamma}_{\gamma\gamma\pi\pi}^{\mu\nu} &=&2ie^2g^{\mu\nu}F_\pi(q_1^2)F_{\pi}(q_2^2),
\end{eqnarray}
where $\boldsymbol{\pi}=(\pi_1,\pi_2,\pi_3)$, $\pi^{\pm}=\frac{\sqrt{2}}{2}(\pi_1\pm i\pi_2)$, $\pi^0=\pi_3$, $D_\mu=\partial_\mu+ieA_\mu$  and $\overline{q}, q_{1,2}$ the momentums of photons. For the effective vertex $\Gamma_{\gamma NN}^{\mu}$, in principle we should take it as that used in Eq. (\ref{eq: Vertex-ND-gamma}), while in the practice, such choice of the effective vertex leads to too complex calculation in the two-loop diagrams. And we approximate it as following  when discuss the TPE contribution from the $\sigma$ meson \cite{zhouhq2016},
\begin{eqnarray}
\Gamma_{\gamma NN}^{\mu} \approx  \widetilde{\Gamma}_{\gamma NN}^{\mu} = ie \gamma^{\mu} F_N(\overline{q}^2).
\end{eqnarray}

\begin{figure}[htbp]
\center{\epsfxsize 2.0 truein\epsfbox{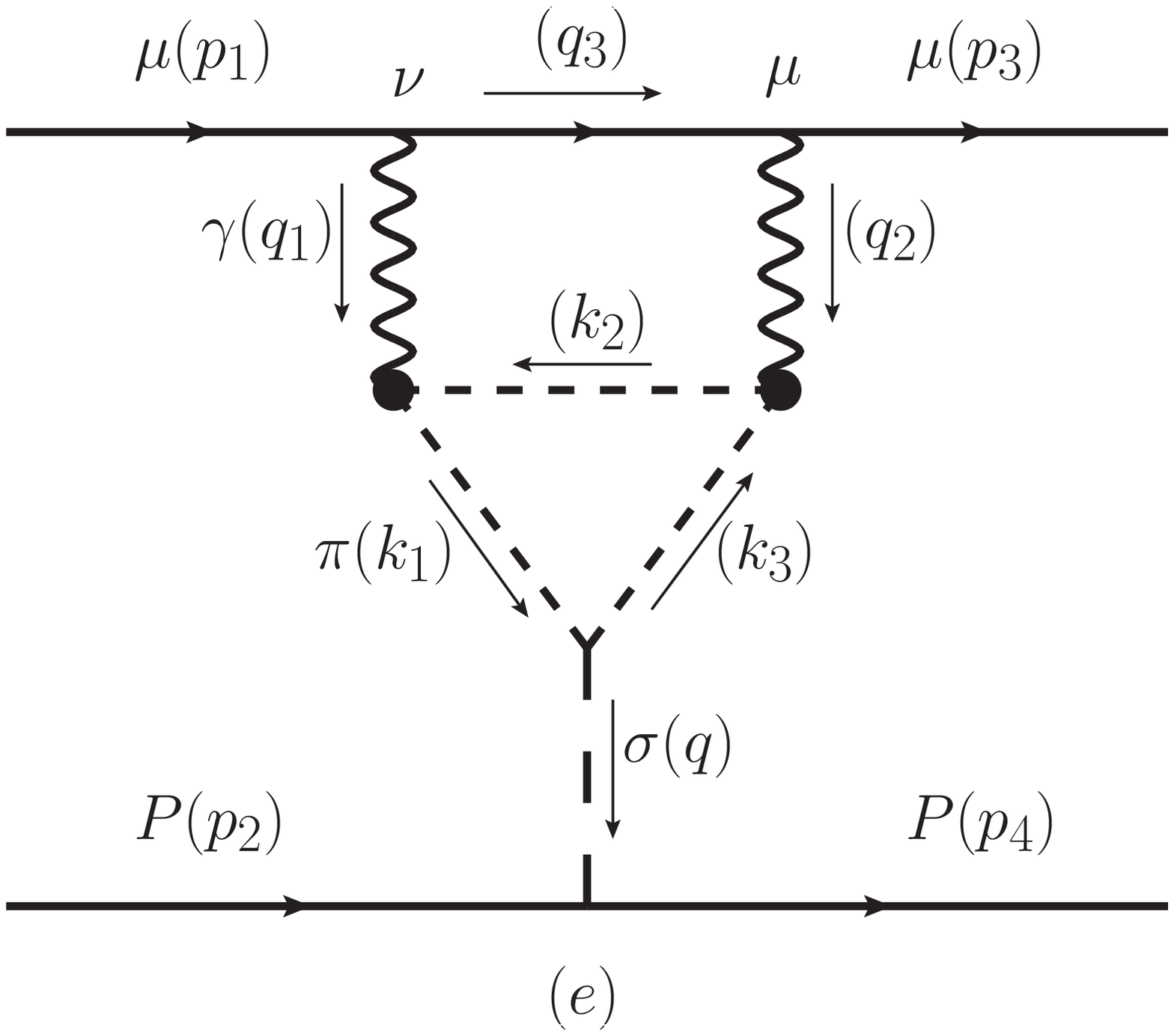}\epsfxsize 2.0 truein\epsfbox{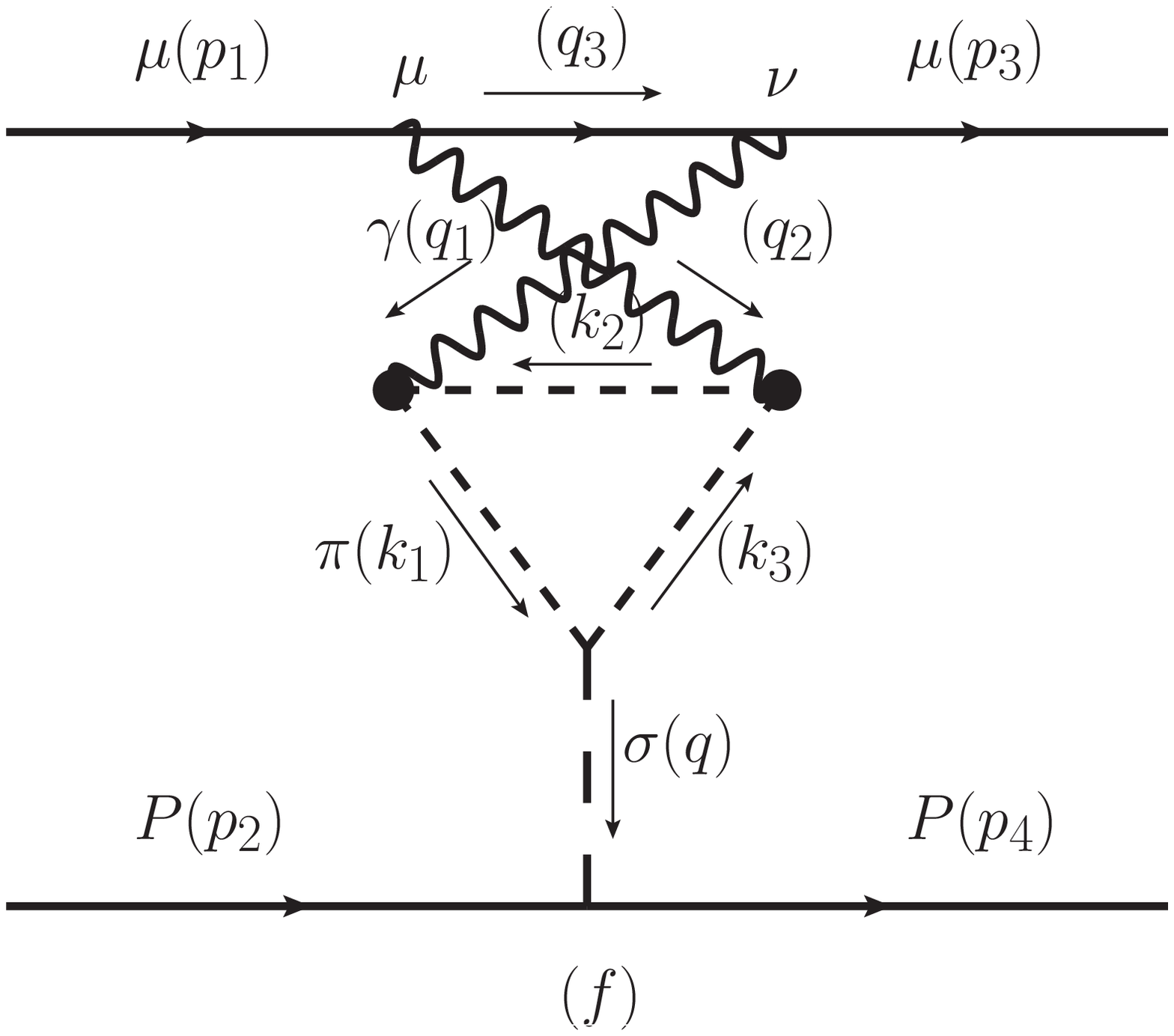}\epsfxsize 2.0 truein\epsfbox{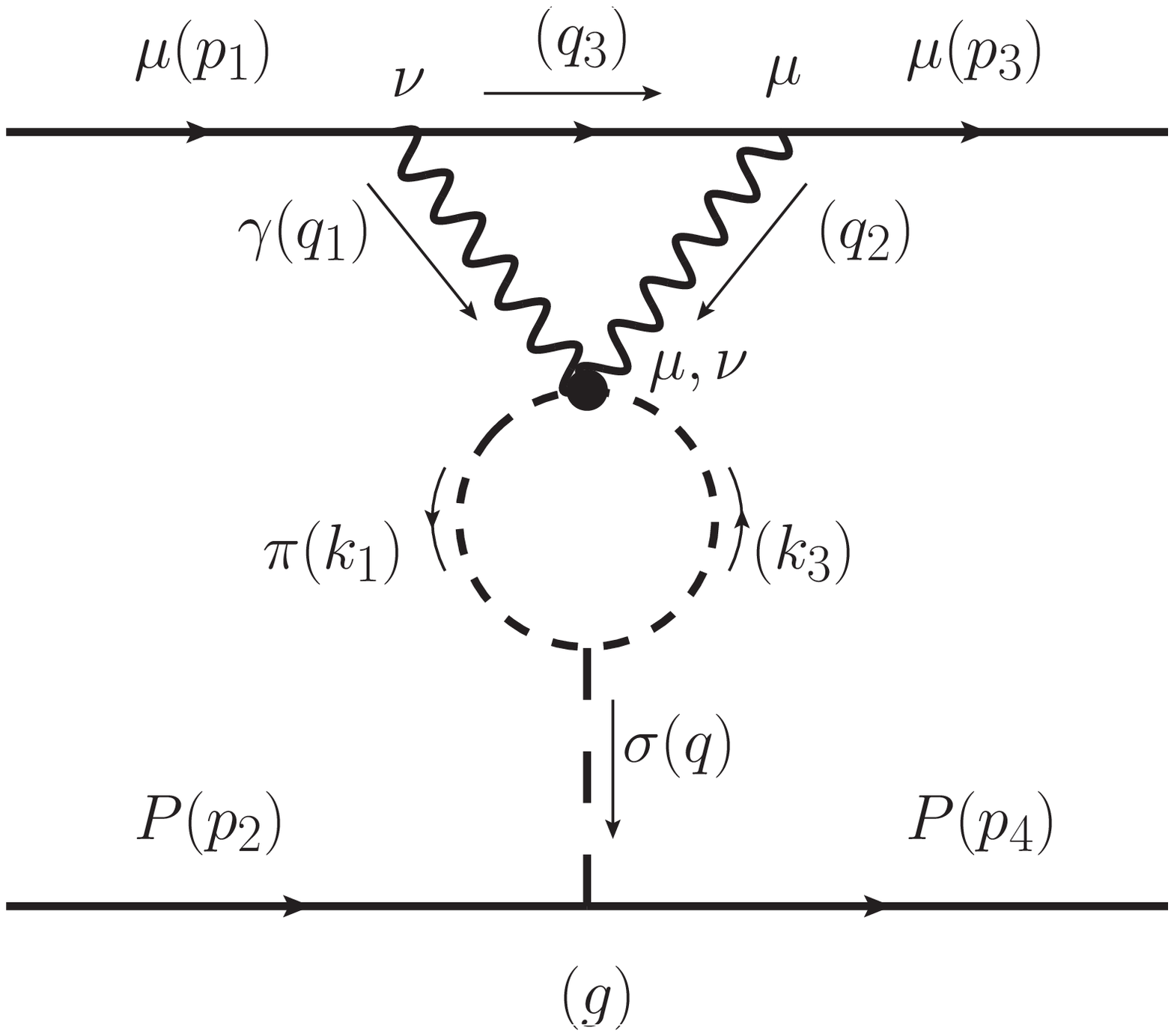}}
\caption{ $\sigma$ meson exchange between muon and proton by  photon and pion loop, (a) box like diagram; (b) crossed-box like diagram; (c) contact like diagram.}
\label{figure:sigma-exchange-diagram-pion}
\end{figure}

\begin{figure}[htbp]
\center{\epsfxsize 2.0 truein\epsfbox{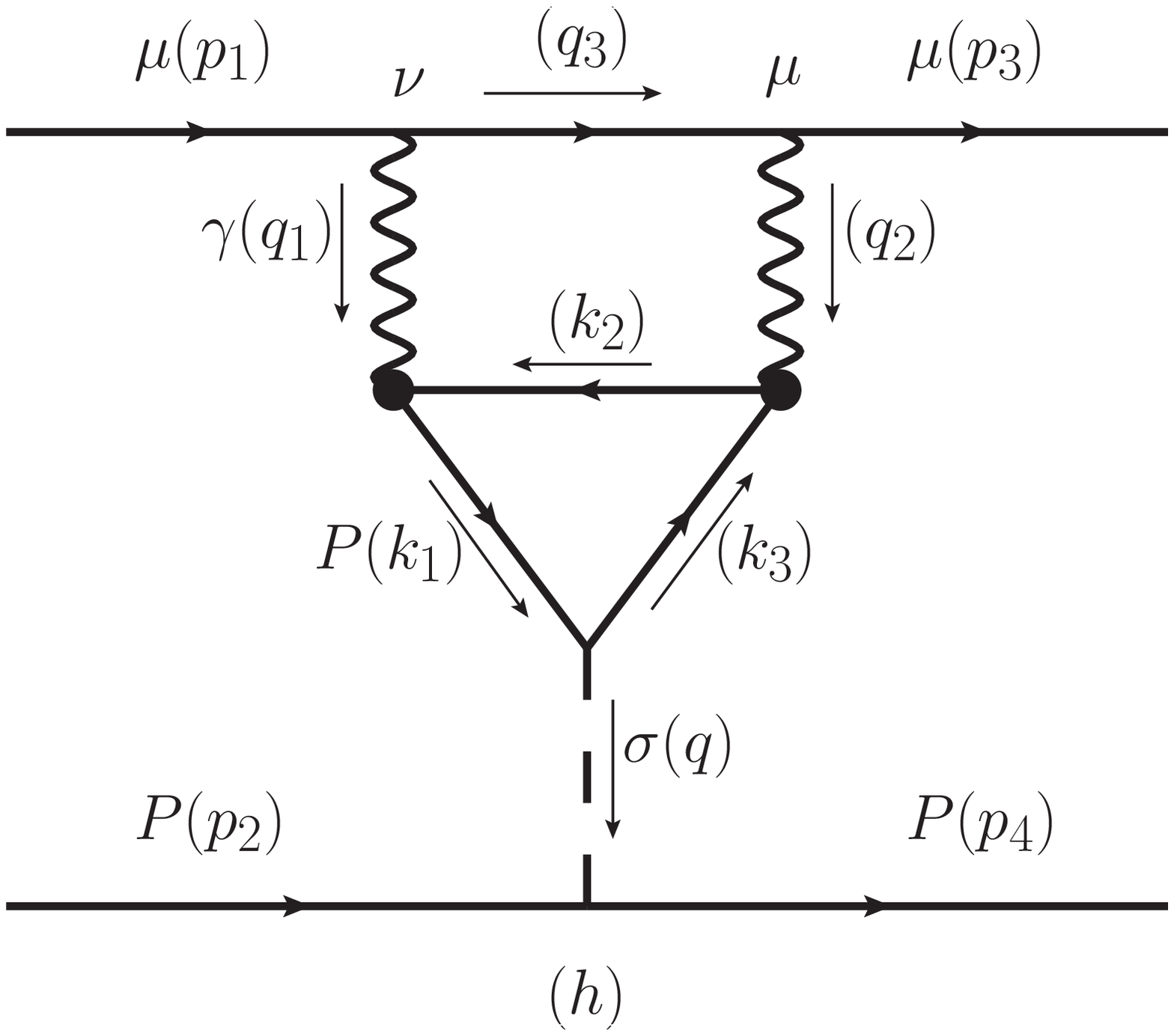}\epsfxsize 2.0 truein\epsfbox{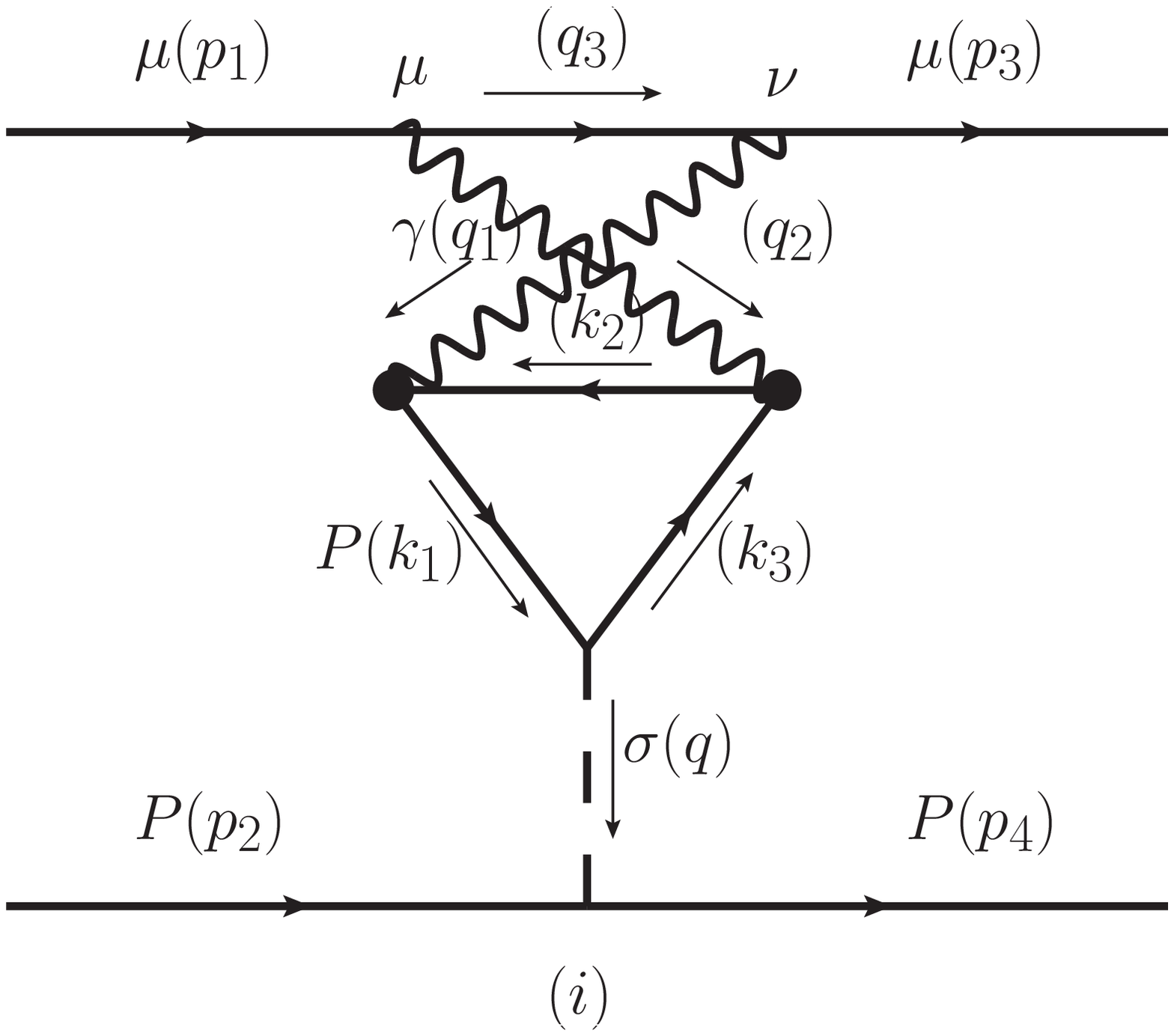}}
\caption{$\sigma$ meson exchange between muon and proton by  photon and proton loop, (d) box like diagram; (e) crossed-box like diagram.}
\label{figure:sigma-exchange-diagram-proton}
\end{figure}

By these effective interactions, the corresponding TPE amplitudes can be written down from the diagrams showed in Fig.  \ref{figure:sigma-exchange-diagram-pion} and Fig. \ref{figure:sigma-exchange-diagram-proton},
\begin{eqnarray}
i\mathcal{M}_{\mu p\rightarrow \mu p}^{(j)}= (i\mathcal{M}_{\mu\rightarrow \mu\sigma^{*}}^{(j)}) \frac{i}{q^2-m_\sigma^2+i\epsilon} (i\mathcal{M}_{p \sigma^{*}\rightarrow p}),
\end{eqnarray}
where $j=(e,f,g,h,i)$, $m_\sigma$ is the mass of $\sigma$ meson and
\begin{eqnarray}
i\mathcal{M}_{p \sigma^{*}\rightarrow p} &=& \overline{u}(p_4,m_N)\widetilde{\Gamma}_{\sigma NN} u(p_2,m_N),
\end{eqnarray}
and in Feynman gauge,
\begin{eqnarray}
i\mathcal{M}_{\mu\rightarrow \mu\sigma^{*}}^{(e)} &=&  \int \frac{d^4k_1 d^4k_2}{(2\pi)^8}\overline{u}(p_3,m_\mu)(-ie\gamma_{\mu}) S_l(q_3)(-ie\gamma_{\nu})u(p_1,m_\mu)S_\gamma(q_1)S_\gamma(q_2) \nonumber \\
&&S_\pi(k_1)S_\pi(k_2)S_\pi(k_3)\widetilde{\Gamma}_{\gamma\pi\pi }^{\mu}(q_1) \widetilde{\Gamma}_{\gamma\pi\pi}^{\nu}(q_2)\widetilde{ \Gamma}_{\sigma \pi\pi}, \nonumber \\
i\mathcal{M}_{\mu\rightarrow \mu\sigma^{*}}^{(f)} &=&  \int \frac{d^4k_1 d^4k_2}{(2\pi)^8}\overline{u}(p_3,m_\mu)(-ie\gamma_{\nu}) S_l(q_3)(-ie\gamma_{\mu})u(p_1,m_\mu)S_\gamma(q_1)S_\gamma(q_2) \nonumber \\
&&S_\pi(k_1)S_\pi(k_2)S_\pi(k_3) \widetilde{\Gamma}_{\gamma\pi\pi }^{\mu}(q_1) \widetilde{\Gamma}_{\gamma\pi\pi}^{\nu}(q_2)\widetilde{ \Gamma}_{\sigma \pi\pi}, \nonumber \\
i\mathcal{M}_{\mu\rightarrow \mu\sigma^{*}}^{(g)} &=&  \int \frac{d^4k_1 d^4k_2}{(2\pi)^8}\overline{u}(p_3,m_\mu)(-ie\gamma_{\mu}) S_l(q_3)(-ie\gamma_{\nu})u(p_1,m_\mu)S_\gamma(q_1)S_\gamma(q_2) \nonumber \\
&&S_\pi(k_1)S_\pi(k_2)\widetilde{\Gamma}_{\gamma\gamma\pi\pi }^{\mu\nu} \widetilde{\Gamma}_{\sigma \pi\pi}, \nonumber \\
i\mathcal{M}_{l\rightarrow l\sigma^{*}}^{(h)} &=&  \int \frac{d^4k_1 d^4k_2}{(2\pi)^8}\overline{u}(p_3,m_{\mu})(-ie\gamma_{\mu}) S_l(q_3)(-ie\gamma_{\nu})u(p_1,m_{\mu})S_\gamma(q_1)S_\gamma(q_2) \nonumber \\
&&(-1)\textrm{Tr}[S_N(k_1)\widetilde{\Gamma}_{\gamma NN}^{\nu}(q_1)S_N(k_2)\widetilde{\Gamma}_{\gamma NN}^{\mu}(q_2)S_N(k_3) \widetilde{\Gamma}_{\sigma NN}], \nonumber \\
i\mathcal{M}_{l\rightarrow l\sigma^{*}}^{(i)} &=&  \int \frac{d^4k_1 d^4k_2}{(2\pi)^8}\overline{u}(p_3,m_\mu)(-ie\gamma_{\nu}) S_l(q_3)(-ie\gamma_{\mu})u(p_1,m_\mu)S_\gamma(q_1)S_\gamma(q_2) \nonumber \\
&&(-1)Tr[S_N(k_1)\widetilde{\Gamma}_{\gamma NN}^{\nu}(q_1)S_N(k_2)\widetilde{\Gamma}_{\gamma NN}^{\mu}(q_2)S_N(k_3) \widetilde{\Gamma}_{\sigma NN}],
\label{eq:amplitudes-proton-loop}
\end{eqnarray}
with
\begin{eqnarray}
S_\pi(\overline{k})=\frac{i}{\overline{k}^2-m_\pi^2+i\epsilon},
\end{eqnarray}
where $m_\pi$ is the mass of pion, $q_{1,2,3}$ and $k_{1,2}$ are the corresponding momentums of the photons, pions and protons showed in the corresponding diagrams of Fig. \ref{figure:sigma-exchange-diagram-pion} and Fig. \ref{figure:sigma-exchange-diagram-proton}.

For comparison, we also define the following effective couplings,
\begin{eqnarray}
i\mathcal{M}_{\mu\rightarrow \mu\sigma^{*}}^{(e+f+g)} &\equiv& \overline{u}(p_3,m_\mu)(-ig_{\sigma \mu\mu}^{(\pi)}) u(p_1,m_\mu), \nonumber \\
i\mathcal{M}_{\mu\rightarrow \mu\sigma^{*}}^{(h+i)} &\equiv& \overline{u}(p_3,m_\mu)(-ig_{\sigma \mu\mu}^{(N)}) u(p_1,m_\mu),
\end{eqnarray}
and these effective couplings $g_{\sigma \mu\mu}^{(\pi,N)}$ can be compared directly with the $f_s$ defined in Ref. \cite{Koshchii2016}.

\section{The input parameters}

\subsection{Input parameters for $N,\Delta$ intermediate states in $s,u$-channel}
For the form factors $F_{1,2}$ in the vertex $\Gamma_{\gamma NN}^{\mu}$, we take the following form as Ref. \cite{Blunden05},
\begin{eqnarray}
F_{1,2}(\overline{q}^2) &=& \sum_{i=1}^3 { n_i \over d_i - \overline{q}^2 }\ ,
\label{eq:N-FFs}
\end{eqnarray}
where the parameters $n_i$ and $d_i$ for the $F_1$ and $F_2$
form factors of the proton can be found in Table~I of Ref. \cite{Blunden05}. Comparing with the calculation in Ref. \cite{Dian-Yong-Chen2013,Tomalak2014} , the choice of the form factors is improved.

The $\Delta$ form factors are taken  as that used in Ref. \cite{zhouhq2015},
\begin{eqnarray}
F^{(1)}_{\Delta}&=&F^{(2)}_{\Delta}=\left(\frac{-\Lambda_{1}^2}{\overline{q}^2-\Lambda_{1}^{2}}\right)^{2}
\frac{-\Lambda_{3}^2}{\overline{q}^2-\Lambda_{3}^{2}},\,\nonumber\\
F^{(3)}_{\Delta}&=&\left(\frac{-\Lambda_{1}^2}{\overline{q}^2-\Lambda_{1}^{2}}\right)^{2}\frac{-\Lambda_{3}^2}{\overline{q}^2-\Lambda_{3}^{2}}
\left [a \frac{-\Lambda_{2}^2}{\overline{q}^2-\Lambda_{2}^{2}}+(1-a)
\frac{-\Lambda_{4}^2}{\overline{q}^2-\Lambda_{4}^{2}}\right ],
\label{D3}
\end{eqnarray}
with $\Lambda_{1}= 0.84 \, $GeV,$\,\Lambda_{2} = 2\,\,$ GeV$,\,
\Lambda_{3} = \sqrt{2}\,\, $GeV, $\Lambda_{4}$ = $0.2$ GeV,
$a = -0.3.$  And the other parameters are taken as
$(g_1, g_2, g_3)=(6.59, 9.08, 7.12)$.  The detail of such choice can be found in Ref. \cite{zhouhq2015}.

\subsection{Input parameters for $\sigma$ intermediate state in $t$-channel}

For the form factor of pion, we  simplify take it as $F_\pi(\overline{q}^2)=-\Lambda^2/(\overline{q}^2-\Lambda^2)$ with $\Lambda=0.77$GeV \cite{Pion-FormFactor}, for $F_N$  for simplify we also take $F_N(\overline{q}^2) = F_\pi(\overline{q}^2)$.

For $g_{\sigma NN}$ and $m_\sigma$, their values can be found in many literatures on the nucleon-nucleon potential, and we list some of these \cite{NN-Boson-1,NN-Boson-2,NN-Boson-3,NN-Boson-4} in the Tab. \ref{Table: values-gsigmaNN-msigmaNN}, where we see there is about $20\%$ difference between the values for $g_{\sigma NN}$ and $m_\sigma$. For simplicity, we take the values in Ref. \cite{{NN-Boson-1}} for our estimation. We also want to point out that the value of $m_\sigma$ can be different with the pole mass of $\sigma$, and it should be understood as the effective or running mass of $\sigma$ in the $t$-channel.

\begin{table}[htbp]
\begin{tabular}
{|p{80pt}<{\centering}|p{60pt}<{\centering}|p{80pt}<{\centering}|p{60pt}<{\centering}|p{90pt}<{\centering}|}
\hline
& $m_\sigma$ (GeV) & $g_{\sigma NN}(Q^2)$ & $\Lambda_\sigma $ (GeV)& $g_{\sigma NN}^2(Q^2=0)/m_\sigma  ^2$ (GeV$^{-2}$) \\
\hline
Ref. \cite{NN-Boson-1}(Tab.5)&0.550&10.20$\frac{\Lambda_\sigma^2-m_\sigma ^2}{\Lambda_\sigma^2+Q^2}$&2.0&294 \\
\hline
Ref. \cite{NN-Boson-2}&0.650&12.78$\frac{\Lambda_\sigma^2-m_\sigma ^2}{\Lambda_\sigma^2+Q^2}$&1.7&282 \\
\hline
Ref. \cite{NN-Boson-3}&0.5325&10.581$\frac{\Lambda_\sigma^2-m_\sigma ^2}{\Lambda_\sigma^2+Q^2}$&2&356 \\
\hline
Ref. \cite{NN-Boson-4}&0.65&13.85$\frac{\Lambda_\sigma^2-m_\sigma ^2}{\Lambda_\sigma^2+Q^2}$&1.8&343 \\
\hline
\end{tabular}
\caption{Values of $m_\sigma$ and $g_{\sigma NN}$ in the literatures.} \label{Table: values-gsigmaNN-msigmaNN}
\end{table}

For $g_{\sigma \pi \pi}$, we take its form as $g_{\sigma \pi \pi} (Q^2) =\widetilde{g}_{\sigma \pi \pi}\frac{\Lambda_\sigma^2-m_\sigma ^2}{\Lambda_\sigma^2+Q^2}$ and match $g_{\sigma \pi \pi}(0)$ with B$\chi$PT \cite{Alarcon2014} by $g_{\sigma \pi\pi}(0) g_{\sigma NN}(0)/m_\sigma ^2=g_A^2 m_N/f_\pi^2\approx 177$ GeV$^{-1}$, which gives $\widetilde{g}_{\sigma\pi\pi}=6.14$ GeV.

\section{Numerical results}

We use the package FeynCalc \cite{FeynCalc9} to deal with the analytical part of the calculation,  use LoopTools \cite{Looptools} to do the numerical integration for one loop diagrams and use FIESTA4 \cite{FIESTA} to do the numerical integration for the two-loop diagrams.

\subsection{Numerical results for TPE corrections from $N$ intermediate state}
Using the expression of the amplitudes, we can get the corresponding cross sections directly as
\begin{eqnarray}
\sigma^{1\gamma}_{\mu p} &=& C_{\mu p} \sum |\mathcal{M}^{(1\gamma)}_{\mu p}|^2, \nonumber \\
\sigma^{1\gamma+2\gamma(N)}_{\mu p} &\equiv& C_{\mu p} \sum[|\mathcal{M}^{(1\gamma)}_{\mu p}|^2+2Re[\mathcal{M}^{(1\gamma)*}_{\mu p}(\mathcal{M}^{(a+b)}_{\mu p}-\mathcal{M}^{(MT)}_{IR,\mu p})]] \nonumber \\
&\equiv& \sigma^{1\gamma}_{\mu p}[1+\delta^{(N,Full)}_{\mu p}-\delta^{(MT)}_{IR,\mu p}] \nonumber \\
&\equiv& \sigma^{1\gamma}_{\mu p}[1+\delta^{(N)}_{\mu p}],
\end{eqnarray}
where $C_{\mu p}$ is a global factor related with the phase space, $\mathcal{M}^{(MT)}_{IR,\mu p}$ refers to the IR part of the amplitudes separated  by the Mao and Tsai's method \cite{Mo-Tasi-1969}, $\delta_{IR,\mu p}^{(MT)}$ is the corresponding correction to the cross section and its explicit expressions can be found in Ref. \cite {Blunden03}.

\begin{figure}[htbp]
\center{\epsfxsize 3.0 truein\epsfbox{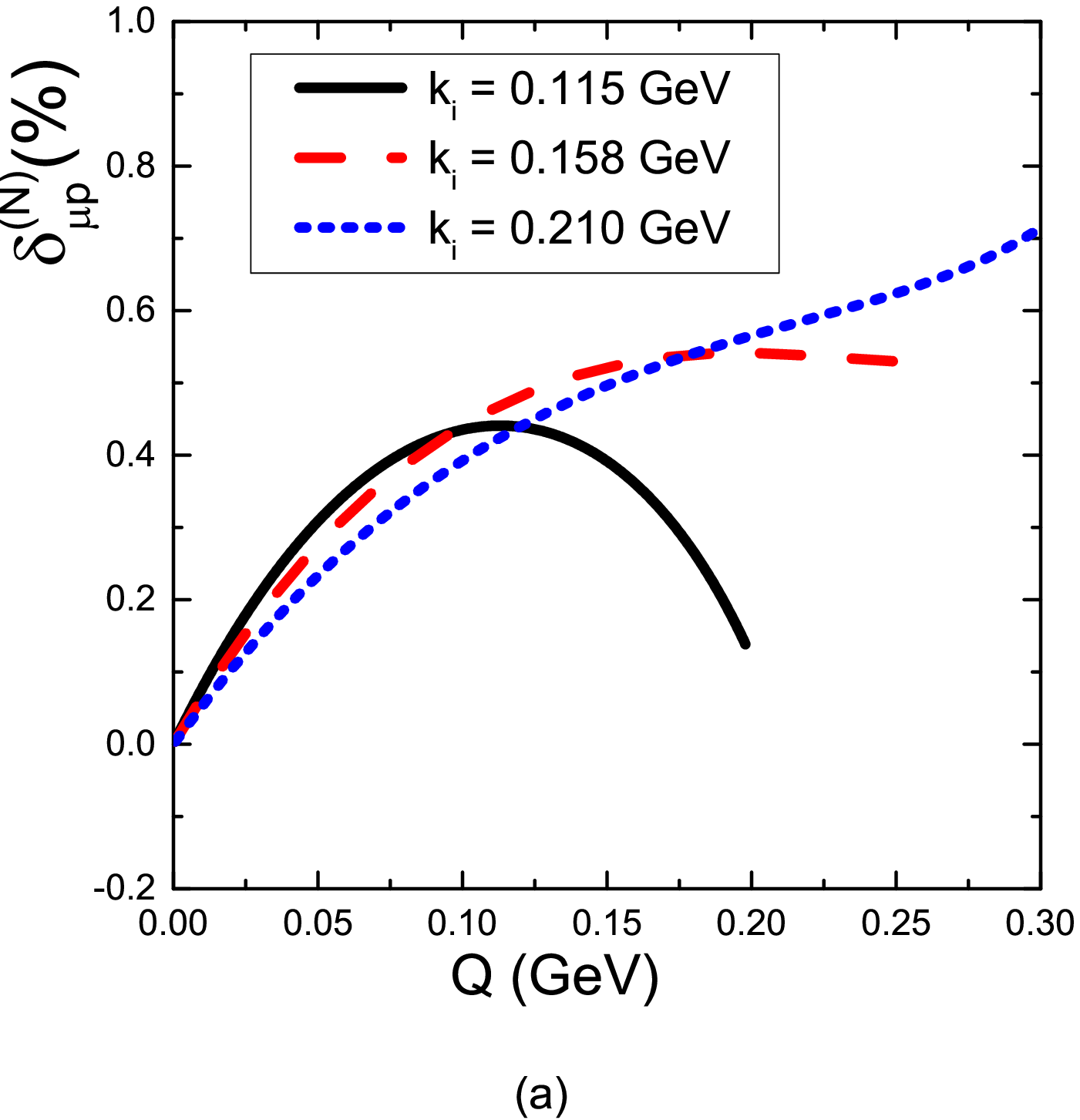}\epsfxsize 3.0 truein\epsfbox{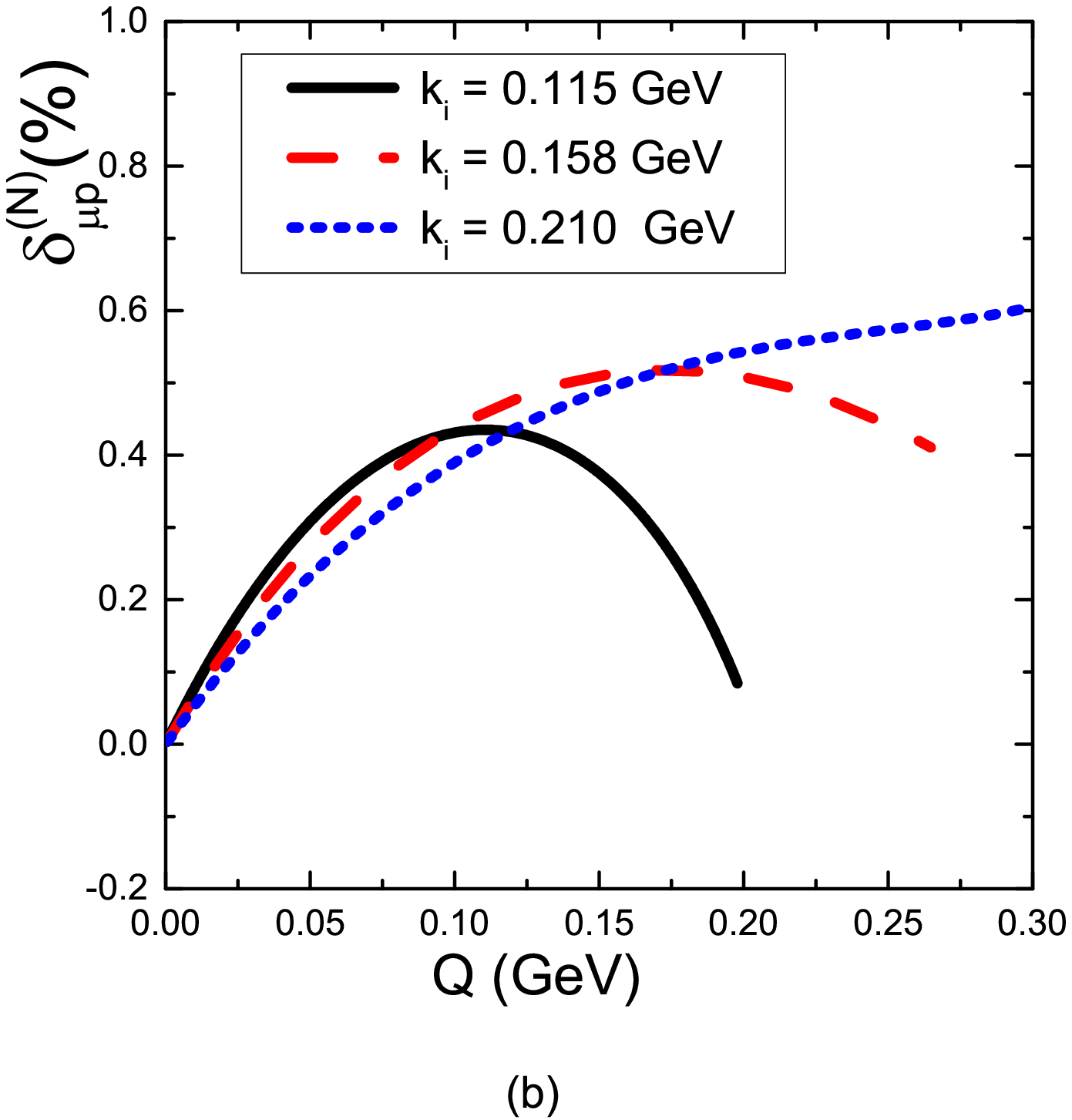}}
\caption{TPE corrections from $N$ intermediate state $\delta^{(N)}_{\mu p}~vs.~Q$ at $k_i=0.115,0.158,0.21$ GeV with $k_i$ the three momentum of the initial muon in the Lab frame. (a) The corrections using Eq. (\ref{eq:N-FFs}) as input; (b) the corrections using Eq. (18) of Ref. \cite{Alarcon2014} as input. }
\label{figure:TPE-N}
\end{figure}

The numerical results for $\delta^{(N)}_{\mu p} vs.~ Q$ at fixed $k_i$ are present in Fig. \ref{figure:TPE-N}(a) where $k_i$ is the magnitude of the three momentum of the initial muon in the Lab frame. Here we use $Q$ but not $Q^2$ as x-coordinate due to the advantage in the following fitting. Also we should note that when the $k_i$ is fixed, there is a maximum value for the $Q$.

Furthermore, we fit the TPE corrections $\delta^{(N)}_{\mu p}$  at the small $k_i$ and $Q$ by the following naive formula,
\begin{eqnarray}
\delta^{(N)}_{\mu p}(Q^2,k_i)& = &[c_{1,\mu p}^{(N)}+c_{2,\mu p}^{(N)}k_i+c_{3,\mu p}^{(N)}k_i^2]Q+[c_{4,\mu p}^{(N)}+c_{5,\mu p}^{(N)}k_i+c_{6,\mu p}^{(N)}/k_i]Q^2 \nonumber \\
&& +[c_{7,\mu p}^{(N)}+c_{8,\mu p}^{(N)}k_i+c_{9,\mu p}^{(N)}k_i^2]Q^3.
\label{eq:TPE-N-fit}
\end{eqnarray}

The numerical results for the fitted parameters are listed in Tab. \ref{Table-c-N}. By these parameters, the $\delta^{(N)}_{\mu p}$ in the full region with $k_i \subseteq ([0.01,0.3]$ GeV and $Q\leq0.4$GeV can be well reproduced and this formula can be used directly to estimate the TPE correction from the $N$ intermediate state in the above momentum region within our model.

\begin{table}[htbp]
\center{
\begin{tabular}
{|p{41pt}<{\centering}|p{41pt}<{\centering}|p{41pt}<{\centering}|p{41pt}<{\centering}|p{41pt}<{\centering}|p{41pt}<{\centering}|}
\hline $c_{1,\mu p}^{(N)}$ &15.2205& $c_{4,\mu p}^{(N)}$ &52.5231 & $c_{7,\mu p}^{(N)}$ &91.8465  \\
\hline $c_{2,\mu p}^{(N)}$ &-70.787& $c_{5,\mu p}^{(N)}$ &-113.801& $c_{8,\mu p}^{(N)}$ &-416.08 \\
\hline $c_{3,\mu p}^{(N)}$ &118.222& $c_{6,\mu p}^{(N)}$ &-10.1527& $c_{9,\mu p}^{(N)}$ &592.395  \\
\hline
\end{tabular}}
\caption{Numerical results for the parameters $c_i^{(N)}$, and the units for both $k_i$ and $Q$ are GeV in the fitting to get $c_i^{(N)}$.}
\label{Table-c-N}
\end{table}

And for comparison, in Fig.  \ref{figure:TPE-N}(b) we also present the numerical results using the form factors Eq.(18) of Ref. \cite{Tomalak2014} as input. Our numerical results are same with that given in Ref. \cite{Tomalak2014} when $\epsilon<1$, while we find there is a minus difference  when $\epsilon>1$, where the definition of $\epsilon$ can be found in Ref. \cite{Tomalak2014}.

\subsection{Numerical results for TPE corrections from $\Delta$ intermediate state}
\begin{figure}[htbp]
\center{\epsfxsize 4.0 truein\epsfbox{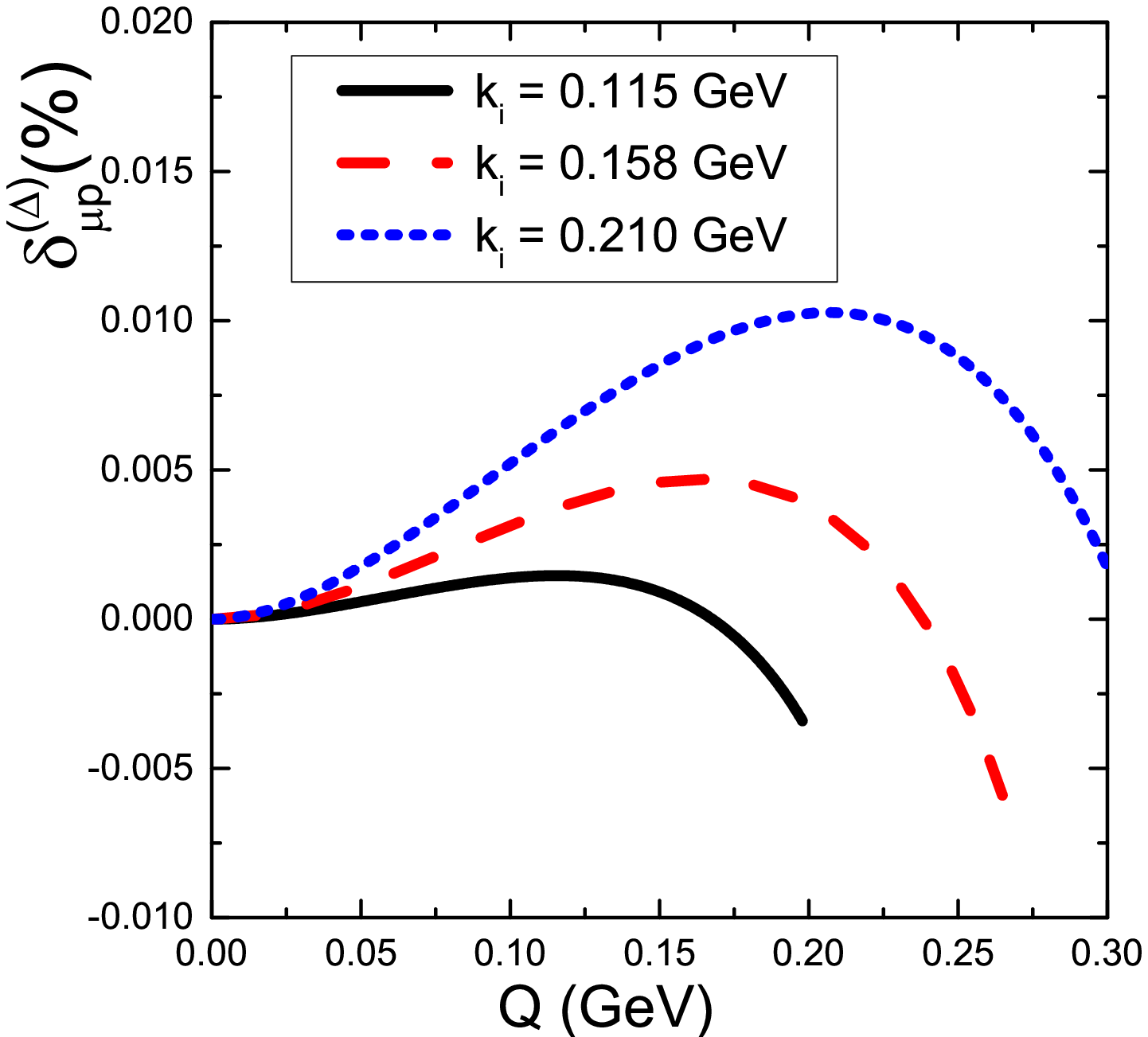}}
\caption{TPE corrections from $\Delta$ intermediate state $\delta^{(\Delta)}_{\mu p}~vs.~Q$ at $k_i=0.115,0.158,0.21$GeV with $k_i$ the three momentum of the initial muon in the Lab frame.}
\label{figure:TPE-D-our-FFs}
\end{figure}

Similar with the $N$ case, we define
\begin{eqnarray}
\sigma^{1\gamma+2\gamma(\Delta)}_{\mu p} &\equiv& C_{\mu p} \sum[|\mathcal{M}^{1\gamma}_{\mu p}|^2+2Re[\mathcal{M}^{1\gamma*}_{\mu p}\mathcal{M}^{(c+d)}_{\mu p}] \nonumber \\
&\equiv& \sigma^{1\gamma}_{\mu p}[1+\delta^{(\Delta)}_{\mu p}].
\end{eqnarray}
The numerical results for the $\delta^{(\Delta)}_{\mu p}$ are presented in Fig. \ref{figure:TPE-D-our-FFs}.

Similarly we fit the $\delta^{(\Delta)}_{\mu p}$ at the small $k_i$ and $Q^2$  as
\begin{eqnarray}
\delta^{(\Delta)}_{\mu p}(Q^2,k_e) &=&[c_{1,\mu p}^{(\Delta)}k_i+c_{2,\mu p}^{(\Delta)}k_i^2+c_{3,\mu p}^{(\Delta)}k_i^3]Q+[c_{4,\mu p}^{(\Delta)}k_i+c_{5,\mu p}^{(\Delta)}k_i^2+c_{6,\mu p}^{(\Delta)}k_i^3]Q^2 \nonumber \\
&&~+[c_{7,\mu p}^{(\Delta)}k_i+c_{8,\mu p}^{(\Delta)}k_i^2+c_{9,\mu p}^{(\Delta)}k_i^3]Q^3.
\end{eqnarray}

\begin{table}[htbp]
\center{
\begin{tabular}
{|p{41pt}<{\centering}|p{41pt}<{\centering}|p{41pt}<{\centering}|p{41pt}<{\centering}|p{41pt}<{\centering}|p{41pt}<{\centering}|}
\hline $c_{1,\mu p}^{(\Delta)}$ & -0.1314& $c_{4,\mu p}^{(\Delta)}$ &0.3633 & $c_{7,\mu p}^{(\Delta)}$ &-19.2295  \\
\hline $c_{2,\mu p}^{(\Delta)}$ &1.0377& $c_{5,\mu p}^{(\Delta)}$ &28.2938& $c_{8,\mu p}^{(\Delta)}$ &36.1717 \\
\hline $c_{3,\mu p}^{(\Delta)}$ & -0.7978& $c_{6,\mu p}^{(\Delta)}$ &-71.6715& $c_{9,\mu p}^{(\Delta)}$ &18.0616  \\
\hline
\end{tabular}}
\caption{Numerical results for the parameters $c_i^{(\Delta)}$, and the units for both $k_i$ and $Q$ are GeV in the fitting to get $c_i^{(\Delta)}$.}
\label{Table-c-D}
\end{table}

The numerical results for the fitted parameters are listed in Tab. \ref{Table-c-D}. The results in the region with $k_i \subseteq [0.1,0.3]$ GeV and $Q\leq0.4$GeV can be well reproduced by this formula and these parameters. The corrections in the region $k_i<0.1$ GeV are almost zero and we do not give a meticulous fitting.

\subsection{Numerical results for TPE corrections from $\sigma$ intermediate state in $t$-channel}

To discuss the TPE corrections from the $\sigma$ meson exchange, we define
\begin{eqnarray}
\sigma^{1\gamma+2\gamma(\sigma,\pi)}_{\mu p} &\equiv& C_{\mu p} \sum[|\mathcal{M}^{1\gamma}_{\mu p}|^2+2Re[\mathcal{M}^{1\gamma*}_{\mu p}\mathcal{M}^{(e+f+g)}_{\mu p}] \nonumber \\
&\equiv& \sigma^{1\gamma}_{\mu p}[1+\delta^{(\sigma,\pi)}_{\mu p}], \nonumber \\
\sigma^{1\gamma+2\gamma(\sigma,N)}_{\mu p} &\equiv& C_{\mu p} \sum[|\mathcal{M}^{1\gamma}_{\mu p}|^2+2Re[M^{1\gamma*}_{\mu p}\mathcal{M}^{(h+i)}_{\mu p}]\nonumber \\
&\equiv& \sigma^{1\gamma}_{\mu p}[1+\delta^{(\sigma,N)}_{\mu p}].
\label{eq:definition-delta-sigma}
\end{eqnarray}

\begin{figure}[htbp]
\center{\epsfxsize 3.3 truein\epsfbox{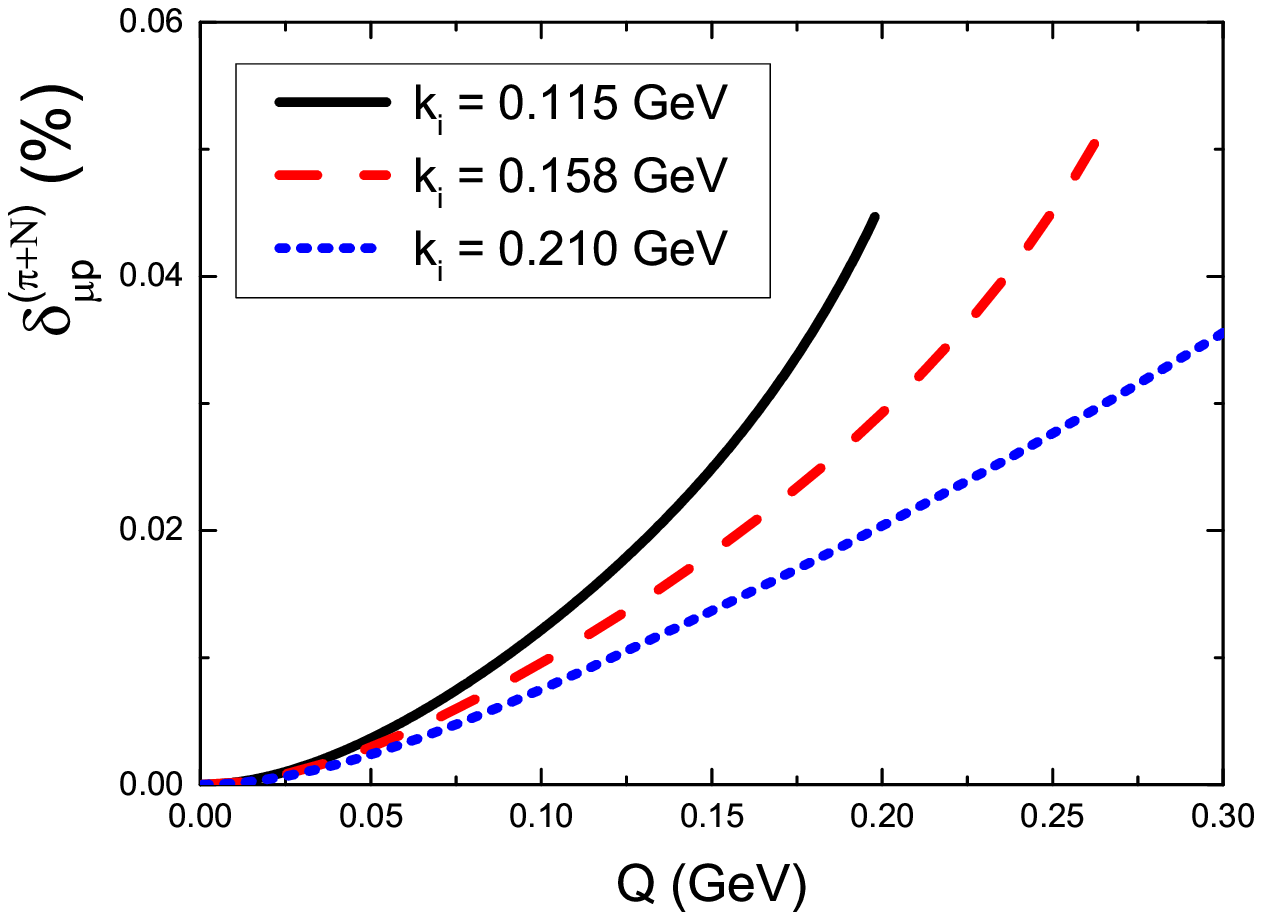}\epsfxsize 3.3 truein\epsfbox{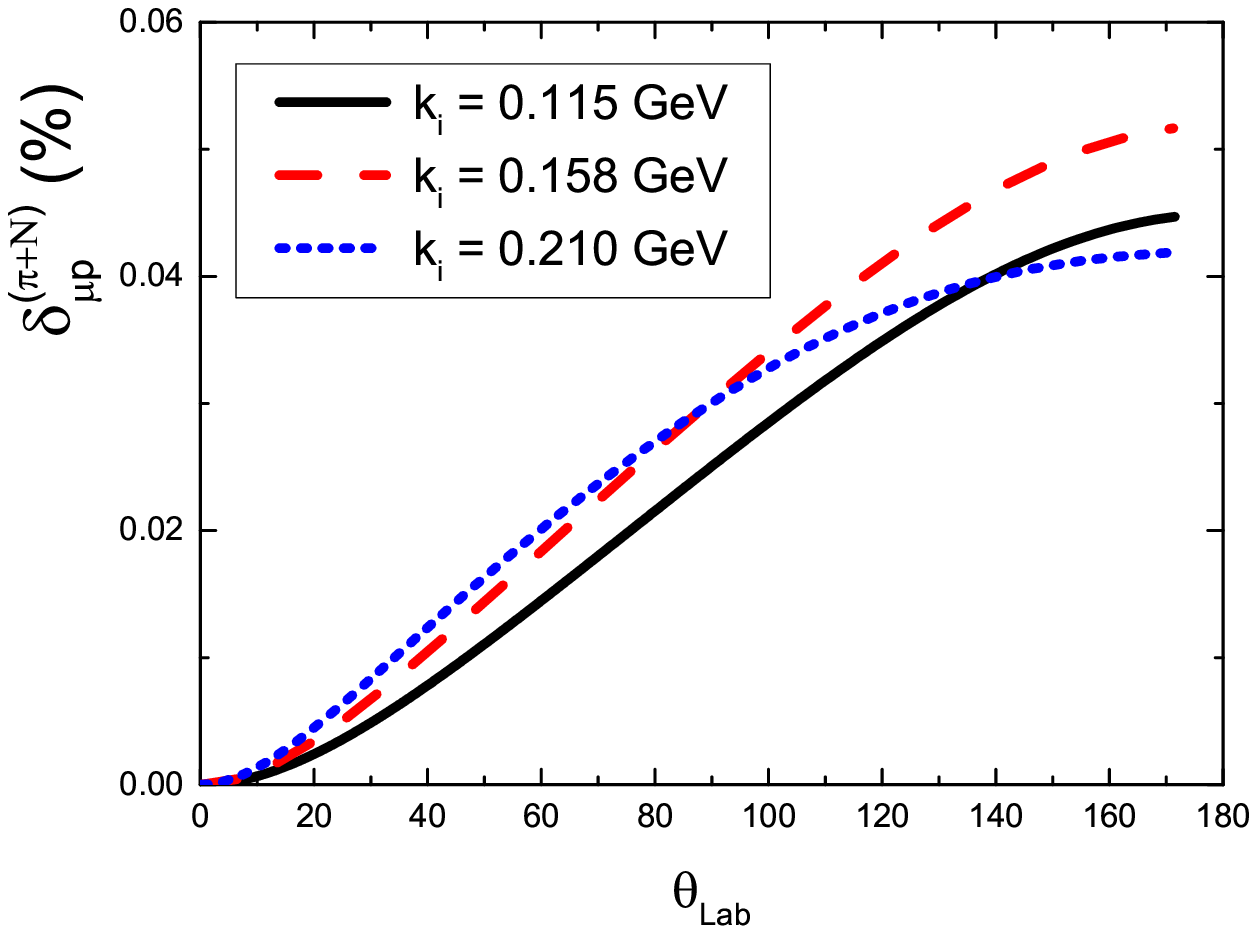}}
\caption{TPE corrections from $\sigma$ intermediate state $\delta^{(\sigma,\pi+N)}_{\mu p}~vs.~Q$ at $k_i=0.115,0.158,0.21$GeV with $k_i$ the three momentum of the initial muon in the Lab frame.}
\label{figure:delta-sigma}
\end{figure}

The numerical results for $\delta^{(\sigma,(\pi+N))}_{\mu p} vs.~Q$ are presented in the left panel of Fig. \ref{figure:delta-sigma}, and  the results $\delta^{(\sigma,(\pi+N))}_{\mu p} vs.~\theta_{Lab}$ which can be compared directly with Fig. 5 of Ref. \cite{Koshchii2016}  are presented in the right panel of Fig. \ref{figure:delta-sigma}, where $\theta_{Lab}$ is the scattering angle of muon in the Lab frame. And we should note that there is a minus difference between our definition of $\delta_{\mu p}^{(\sigma,\pi+N)}$ by Eq. (\ref{eq:definition-delta-sigma}) and that by Eq. (22) of Ref. \cite{Koshchii2016}.

Different with the $N,\Delta$ cases, the dependence of the TPE corrections from the $\sigma$ meson exchange on the effective coupling $g_{\sigma \mu \mu}^{(\pi,N)}$ can be expressed in an explicit form \cite{Koshchii2016}. So we do not fit the dependence of $\delta_{\mu p}^{(\pi+N)}$ on $k_i$ and $Q$, but express the $g_{\sigma \mu \mu}^{(\pi,N)}$ at the small $Q$ as,
\begin{eqnarray}
g_{\sigma \mu\mu}^{(\pi)}(Q^2) &=& [c_{0}^{(\pi)}+ c_{1}^{(\pi)}Q + c_{2}^{(\pi)}Q^2+ c_{3}^{(\pi)}Q^3]g_{\sigma \pi\pi}, \nonumber \\
g_{\sigma \mu\mu}^{(N)}(Q^2) &=& [c_{0}^{(N)}+ c_{1}^{(N)}Q + c_{2}^{(N)}Q^2 + c_{3}^{(N)}Q^3]g_{\sigma NN},
\end{eqnarray}
and we take $c_{1,2}^{(\pi,N)}$ from Ref. \cite{zhouhq2016} where the function SDExpandAsy in FIESTA is used to calculate, and fit $c_{3,4}^{(\pi, N)}$ from the $g_{\sigma \mu \mu}^{(\pi,N)}$ in the region $Q\subseteq [0.01,0.4]$ GeV and at last we have the parameters as Tab. \ref{Table-c-piN}. And by these parameters, the behavior of $g_{\sigma \mu\mu}^{(\pi,N)}$ at $Q\leq0.4$ GeV can be well reproduced. We should note that $c_{0,1,2,3}^{(\pi,N)}$ are only dependent on the masses of muon, pion, proton and the corresponding form factors in $\Gamma_{\gamma\pi\pi}^{\mu}$ and $\Gamma_{\gamma NN}^{\mu}$. And the $\sigma$ related property is included in the factors $g_{\sigma\pi\pi}$ and $g_{\sigma N N}$.

\begin{table}[htbp]
\center{
\begin{tabular}
{|p{45pt}<{\centering}|p{45pt}<{\centering}|p{45pt}<{\centering}|p{45pt}<{\centering}|p{45pt}<{\centering}|}
\hline ~~~&$c_{0}^{(i)}$   &$c_{1}^{(i)}$   &$c_{2}^{(i)}$ &$c_{3}^{(i)}$  \\
\hline $i=\pi$ &5.2770  & -28.7494   & 67.1914  & -64.4362   \\
\hline $i=N$   & 1.0755 & -4.7336 & 12.3169 & -14.7901  \\
\hline
\end{tabular}}
\caption{Numerical results for the parameters $c_j^{(i)}$ with $j=0,1,2,3$ and $i=\pi,N$, and the unit for $Q$ is GeV in the fitting to get $c_j^{(i)}$.}
\label{Table-c-piN}
\end{table}

\begin{figure}[htbp]
\center{\epsfxsize 4.0 truein\epsfbox{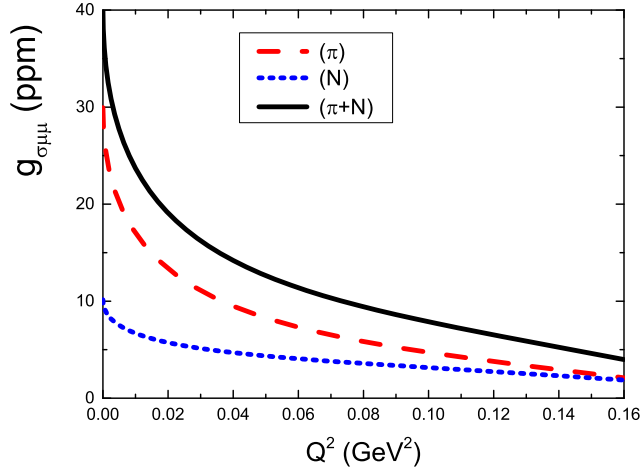}}
\caption{Numerical results for $g_{\sigma\mu\mu}~vs.~ Q^2$ which can be compared directly with the $f_s$ in Ref. \cite{Koshchii2016}  and ppm refers to $10^{-6}$.}
\label{figure:g-sigma-mu-mu}
\end{figure}

To compare with the effective coupling $f_s$ defined in Ref. \cite{Koshchii2016}, we also present the $Q^2$ dependence of $g_{\sigma\mu\mu}^{(\pi,,N,\pi+N)}$ in Fig. \ref{figure:g-sigma-mu-mu}.

\subsection{Discussion and Summary}
The numerical results presented in Fig. \ref{figure:TPE-N} show that the TPE corrections from the $N$ intermediate state are almost independent on the input form factors when $Q<0.2$ GeV and $k_i<0.21$ GeV, this is natural since the different input form factors are almost same at the very low momentum transfer. And when $k_i=0.21$ GeV and $Q=0.3$ GeV, there is sizable difference (about 15\% difference) between our results and that in Ref. \cite{Tomalak2014}, which means the careful choice of the form factors is meaningful when $Q>0.25$ GeV. And the naive formula Eq. (\ref{eq:TPE-N-fit}) can be used directly for other analysis in the region with $k_i \subseteq [0.01,0.3]$ GeV and $Q\leq0.4$GeV.

The corrections from the $\Delta$ intermediate state at the low momentum transfer are much smaller than that from the $N$ intermediate state, and can be neglected when $k_i<0.158$GeV, and even when $k_i=0.21$ GeV and $Q\sim 0.22$ GeV, the correction is about $2\%$ of that from the $N$ intermediate state. Comparing our results with the corrections from the inelastic state estimated by Ref. \cite {Tomalak2016}, we can see that the magnitudes are in the same order, while our results are smaller than theirs. The reason of this difference maybe due to the effects from the $\pi N$ inelastic state and the decay width of $\Delta$. Since in the discussed momentum transfer region this correction is much smaller than that from the $N$ intermediate state, we do not go to discuss this in detail.

For the corrections from the $\sigma$ meson exchange $\delta_{\mu p}^{(\sigma,\pi+N)}$, the general property of our results and those in Ref. \cite{Koshchii2016} is similar when $k_i=0.115,0.158,0.21$ GeV.  For the effective coupling $g_{\sigma \mu \mu}$ we can find that at the small $Q^2$ our results are similar with the results showed in Fig.4 of Ref. \cite{Koshchii2016}, while at $Q^2=0.16$ GeV$^2$, we can find that our results are only about an half of that given in Ref. \cite{Koshchii2016} (shaded region). In other words, $g_{\sigma \mu \mu}$ decreases much quickly in our method than that estimated in Ref. \cite{Koshchii2016}.

In summary, in this work, the TPE corrections to the unpolarized $\mu p$ scattering due to the $N,\Delta$ and the $\sigma$ intermediate states are discussed in the hadronic model. And we find at the small $k_i$ and $Q^2$, the corrections from the $N$ intermediate state are dominant, and the corrections from  the $\Delta$ and the $\sigma$ intermediate states are smaller than $0.05\%$. This property is same with the calculation given in the literatures by other methods. And in our work, the form factors for $\gamma NN$ are improved to estimate the  corrections from the $N$ intermediate state and a naive formula which can well reproduce the corrections in the region with $k_i \subseteq [0.01,0.3]$ GeV and $Q\leq 0.4$GeV is given.

\section{Acknowledgments}
This work is supported by the  National Natural Science Foundations of China
under Grant No. 11375044 and in part by the Fundamental Research Funds for the Central Universities under Grant No. 2242014R30012. The author thanks A.V. Smirnov and Wen-Long Sang for the help on FIESTA, and thank Shin Nan Yang for the helpful suggestion.

\section{Appendix A: Some relations}
In this Appendix, we list the relations between some quantities used in the literatures, and we take $k_i$ and $Q$ as the basic variables.
\begin{eqnarray}
Q^2_{max}&=&\frac{4m_N^2 k_i^2}{2E_im_N+m_N^2+m_\mu^2}, \nonumber \\
cos\theta_{Lab} &=& \frac{2m_N k_i^2-Q^2(E_i+m_N)}{k_i \sqrt{4m_N^2k_i^2-4E_i m_N Q^2+Q^4}}, \nonumber \\
tan \frac{\theta_B}{2} &=& \frac{Q\sqrt{Q^2+4m_N^2}}{2\sqrt{4m_N^2k_i^2-Q^2(2E_im_N+m_N^2+m_\mu^2)}}, \nonumber \\
E_{f} &=& \frac{E_im_N-Q^2}{2m_N},   \nonumber \\
\epsilon &\equiv& \frac{16v^2-Q^2(Q^2+4m_N^2)}{16v^2-Q^2(Q^2+4m_N^2)+2(Q^2+4m_N^2)(Q^2-2m_\mu^2)},
\end{eqnarray}
where $E_i=\sqrt{k_i^2+m_\mu^2}$, $v=m_N(E_i+E_f)/2$, $\theta_{Lab}$ is the scattering angle of finial muon in the Lab frame, $\theta_{B}$ is the scattering angle in the Breit frame, the definition of $\epsilon$ is taken from Ref. \cite{Alarcon2014}. And also we have
\begin{eqnarray}
\frac{1}{4}\sum_{spin} |M^{1\gamma}_{\mu p}|^2&=& e^4(g_1F_1^2+g_2F_2^2+g_3F_1F_2), \nonumber \\
\frac{1}{4}\sum_{spin} 2 Re[M^{1\gamma*}_{\mu p} M^{\sigma}_{\mu p}] &=& e^2g_{\sigma \mu\mu}^{(\pi+N)} g_{\sigma NN} g_4(4F_1m_N^2-F_2Q^2),
\end{eqnarray}
with
\begin{eqnarray}
g_1&=&2(1-\frac{4E_i m_N+2m_N^2+2m_\mu^2}{Q^2}+\frac{8E_i^2m_N^2}{Q^4}), \nonumber \\
g_2&=&1-\frac{2E_i}{m_N}+\frac{4k_i^2}{Q^2},\nonumber \\
g_3&=&4-\frac{8m_\mu}{Q^2},\nonumber \\
g_4&=&\frac{2m_\mu(4E_im_N-Q^2)}{m_NQ^2(m_\sigma^2+Q^2)}.
\end{eqnarray}


\end{document}